\documentclass{article}

\usepackage{PRIMEarxiv}

\usepackage[utf8]{inputenc} 
\usepackage[T1]{fontenc}    
\usepackage{hyperref}       
\usepackage{url}            
\usepackage{booktabs}       
\usepackage{amsfonts}       
\usepackage{nicefrac}       
\usepackage{microtype}      
\usepackage{lipsum}
\usepackage{fancyhdr}       
\usepackage{graphicx}       
\graphicspath{{media/}}     
\usepackage{ptdr-definitions}  
\usepackage{heppennames2}      

\usepackage{amsmath,amsfonts,bm}
\usepackage{ amssymb }

\def\eqref#1{equation~\ref{#1}}

\def\1{\bm{1}}

\def\vc{{\bm{c}}}

\def\vf{{\bm{f}}}
\def\vg{{\bm{g}}}
\def\vh{{\bm{h}}}
\def\vi{{\bm{i}}}

\def\vo{{\bm{o}}}

\def\vx{{\bm{x}}}

\def\mW{{\bm{W}}}

\DeclareMathAlphabet{\mathsfit}{\encodingdefault}{\sfdefault}{m}{sl}
\SetMathAlphabet{\mathsfit}{bold}{\encodingdefault}{\sfdefault}{bx}{n}

\usepackage{mathtools}

\usepackage{graphicx,amsfonts,amscd,amssymb,bm,url,color,latexsym,bbm,amsthm,amsmath}
\usepackage{physics}
\allowdisplaybreaks
\usepackage[capitalize,nameinlink]{cleveref}

\renewcommand{\mathbf}{\boldsymbol}

\newcommand{\bb}{\mathbb}

\newcommand{\reals}{\bb R}

\newcommand{\R}{\reals}

\title{Predicting the Future of the CMS  Detector: Crystal Radiation Damage and Machine Learning at the LHC
}

\author{}

\begin{document}
\maketitle
\vspace{-25mm}
\begin{center}
    \textbf{
    Bhargav Joshi\textsuperscript{2,\dag},
    Taihui Li\textsuperscript{1,\dag},
    Buyun Liang\textsuperscript{1,\dag},
    Roger Rusack\textsuperscript{2,\dag},
    Ju Sun\textsuperscript{1,\dag}
    }
    \\
    \textsuperscript{1} Department of Computer Science \& Engineering, University of Minnesota, Minneapolis, USA
    \\
    \textsuperscript{2} Department of Physics, University of Minnesota, Minneapolis, USA
    \\
    \textsuperscript{\dag} These authors contributed equally to this work 
    \\
    \textit{\{liang664, joshib, lixx5027, rusack, jusun\}@umn.edu}
\end{center}
\vspace{10mm}

\begin{abstract}
  The 75,848 lead tungstate crystals in CMS experiment at the CERN Large Hadron Collider are used to measure the energy of electrons and photons produced in the proton-proton collisions. The optical transparency of the crystals degrades slowly with radiation dose due to the beam-beam collisions. The transparency of each crystal is monitored with a laser monitoring system that tracks changes in the optical properties of the crystals due to radiation from the collision products. Predicting the optical transparency of the crystals, both in the short-term and in the long-term, is a critical task for the CMS experiment. We describe here the public data release, following FAIR principles~\cite{fairmetrics}, of the crystal monitoring data collected by the CMS Collaboration between 2016 and 2018.  Besides describing the dataset and its access, the problems that can be addressed with it are described, as well as an example solution based on a Long Short-Term Memory neural network developed to predict future behavior of the crystals.
\end{abstract}


\section{Introduction}

The Compact Muon Solenoid (CMS) experiment at the CERN Large Hadron Collider (LHC) is designed to detect and measure particles produced in the proton-proton collisions at the LHC~\cite{CMS_2008}. 
One of the components of the CMS detector is an electromagnetic calorimeter (ECAL) made of lead tungstate crystals that is used to measure the energy of electrons and photons produced in the collisions. This was the principle detector used to detect the Higgs boson in the two-photon decay channel in the discovery of the Higgs boson in 2012~\cite{Chatrchyan:2012ufa}, in the precision measurement of the Higgs boson mass~\cite{HGG}, and in many other measurements made by the CMS Collaboration.
Changes in the crystal transparency caused by radiation from the collisions in the LHC lead to changes in the response to electrons and photons that needs to be tracked and corrected for. 
Predicting these changes both in the short term and in the long term is a critical question for the CMS experiment. We describe here the public release following FAIR principles~\cite{Chen_2022} of the crystal monitoring data collected by CMS between 2016 and 2018.  Besides describing the dataset and its access, and the problem to be addressed with it, we provide links to an example solution based on a long short-term memory (LSTM) neural network, developed to predict future behavior of the crystals. 

\subsection{Electromagnetic Calorimeter (ECAL)}

There are 75,848 lead tungstate crystals in the ECAL, of which 61,200 are arranged in a barrel surrounding the interaction point where the beams collide, and the barrel is capped by two endcap calorimeters, each consisting of 7,324 crystals~\cite{ECALTDR}. The crystals in CMS measure approximately 22 \cm in length with a $2\times2 \cm^2$ section and weigh $\approx 1.1 \unit{kg}$. Lead tungstate crystals are optically transparent and emit a short pulse of light when they absorb ionizing radiation.  Due to the intense radiation while the LHC is in operation, the optical transparency of the crystals is reduced over time. 
This reduction is due to the creation by the radiation of atomic-level impurities in the crystal that act as color centers~\cite{914439-xtal-color-centers} that absorb light propagating through the crystal. 
Deep well impurities are stable and persist for years and the radiation damage is permanent, while shallow impurities are metastable and are short-lived. During beam-beam collisions the optical transmission is reduced and it partially recovers when the beams are off as the meta-stable states decay.

When the LHC is operating, the beam-beam collisions are continuous, with a `fill' generally lasting for approximately 18 hours and a four- to six-hour interval between fills. There are longer intervals when there are no beams for maintenance of the LHC and the detectors. These typically last on the order of a day, and every year there are several-month-long shut downs. 

During a fill there are proton-proton collisions at a rate of about 1 GHz at the center of the CMS detector. These collisions produce a copious number of secondary particles that travel through the detector and are absorbed. It is these secondary particles that induce the production of color centers in the ECAL crystals, and the more particles produced in the proton-proton collisions: the more color centers are created in the crystals. Hence the number of color centers produced is proportional to the number of collisions in the LHC, which is measured in terms of the `luminosity'.  Luminosity has units of inverse centimeters, or, since these are sub-nuclear processes, in terms of inverse femtobarns (fb$^{-1}$), where a femtobarn is $1\times10^{-39} \rm{cm}^2$.\ \footnote{For one inverse femtobarn of delivered luminosity we expect about one event with a cross-section of a femtobarn to occur. For example, at the LHC the process where a Higgs boson is produced has a cross section of 55,400 femtobarns, then with one inverse femtobarn of luminosity we expect that 55,400 Higgs bosons will be produced. Unfortunately many fewer are detected.}

Since detailed knowledge of the crystal's light output due to ionizing radiation is essential for the physics measurements, the transparency of each crystal is carefully monitored with a laser monitoring system that injects pulses of light into each crystal at intervals of approximately every 40 minutes. 

The principle of operation of the crystal calorimeter, which is designed to detect electrons and photons with energies of 1\GeV or more, is as follows: When a high energy electron or photon ($>100 \MeV $) is incident on the crystal electron and positron pairs are produced, these in turn interact with the atomic nuclei and radiate (bremsstrahlung) photons. These in turn produce more electron-positron pairs, which in turn radiate photons, though always with less energy. This sets up a cascade where photons and electron-positron pairs are produced that continues until the photons have insufficient energy to create further electron-positron pairs ($< 1 \MeV$). During this process the electrons and positrons produced propagate in the crystal and ionize the atoms causing scintillation light (optical photons) to be produced that can be detected with a photodetector coupled to the crystal. The operating principle of the CMS ECAL is that the amount of ionizing radiation is linearly proportional to the energy of the incident electron or photon, and hence the light output is proportional to the energy of the incident particle. 

The cascade is a stochastic process, thus the energy measured with the crystals is measured with a resolution that can be parameterized as:

\begin{equation}
\dfrac{\sigma_{E}}{E}=\dfrac{S}{\sqrt{E}}\oplus\dfrac{N}{E}\oplus C
\end{equation}

where the first term on the right is the \textit{stochastic} term, which is the contribution to the resolution of natural fluctuations in the cascade; second is the \textit{noise} term due to electronic noise, and the third is the \textit{constant} term that accounts for energy leakage and other signal losses. The parameters $S$, $N$ and $C$ have been measured with a prototype in a dedicated particle beam. For electrons with $p_{T}$ greater than 10 GeV, the energy resolution is better than 1\%. 

\begin{figure}[htbp]
	\centering
	\includegraphics[width=0.8\linewidth]{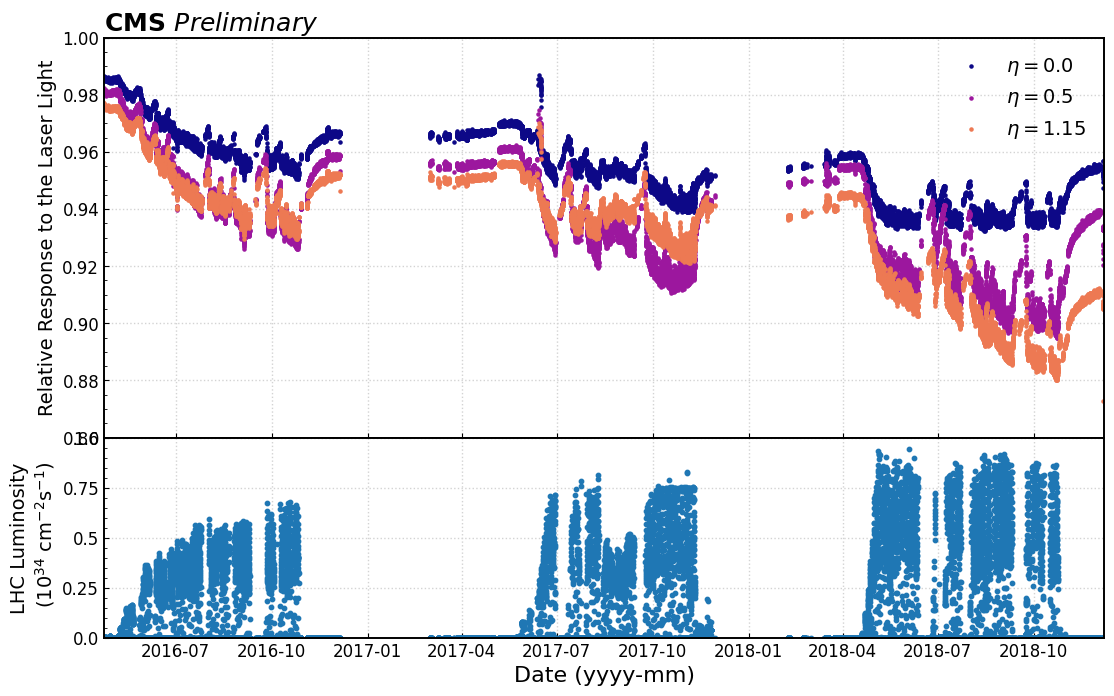}
	\caption{Relative response of crystal to laser light in different $\eta$ regions of ECAL between 2016 and 2018 when the total luminosity of the LHC was 151~fb$^{-1}$. The crystals closest to the center of the  barrel ($|\eta| = 0$)  all have less light loss compared with the crystals that are closest to the end of the barrel  ($|\eta| = 1.15$).}
	\label{fig:crystal_transparency_runII}
\end{figure}

A convenient approximation to the production angle is the pseudorapidity ($\eta$)~\cite{ECALTDR}, which is equal to zero in the center of the barrel and increases to $\pm 1.41$ at the end of the barrel.
The changes in the crystal transparency for different regions as they are being irradiated for the barrel calorimeter can be seen in Figure \ref{fig:crystal_transparency_runII}.

\section{Data Challenge}
The details of the color center formation under radiation are poorly understood. It is thought that they are primarily determined by the atomic-level defects in the crystals when they were first grown and, thus are determined by the individual history of each crystal. From this it follows that the best way to predict the future behavior of a crystal is by examining their past behavior under radiation. 

It is valuable to the CMS collaboration to be able to accurately predict how the ensemble of crystals will respond to future irradiation cycles with damage and recovery. In particular there is interest in predicting the level of light-loss for a group of crystals after a duration of the order one day, when the integrated luminosity is $\approx 1 \rm{fb}^{-1}$. Another point of interest is to predict the performance of the crystals in the barrel ECAL up to the end of operation of the High-Luminosity LHC in 2040, when the total delivered luminosity is expected to be 3000 fb$^{-1}$

In the examples given below the short-term problem is tackled. The long term problem is left as an open challenge.

\section{Datasets}
\label{datasets}
The response to injected laser pulses for every crystal is measured approximately every 40 minutes and stored in an offline database. Data corresponding to the laser response of all 75848 crystals from 2016 through 2018 were extracted from the database.  Each entry has a timestamp corresponding to the time when the measurement was taken, and the current intensity of the beam-beam collisions (the instantaneous luminosity) which is a measure of the radiation dose being received by the crystals.  

The dataset consists of the following elements corresponding to a single measurement.

\begin{itemize}
    \item \textbf{xtal\_id}: Crystal identification number within ECAL ranging from [0, 75848]. 
    \item \textbf{start\_ts}: Start of the Interval of Validity (IOV). An Interval of Validity corresponds to a time during which a measurement is taken for a single crystal. In other words, each IOV contains one measurement per crystal.
    \item \textbf{stop\_ts}: End of the Interval of Validity (IOV).
    \item \textbf{laser\_datetime}: Timestamp of the measurement for a given crystal within an IOV. The timestamp lies between the start of IOV and end of IOV.
    \item \textbf{calibration}: APD/PD ratio taken at laser\_datetime. This value is used to quantify the transparency of the crystal at the time of measurement.
    \item \textbf{time}: Time corresponding to the luminosity measurement (obtained from BRIL) closest to the time when the laser measurement was taken.
    \item \textbf{int\_deliv\_inv\_ub}: Approximate integrated luminosity delivered up to the measurement, in the units of inverse microbarns.
\end{itemize}

To ensure the FAIR-ness of the publication of the dataset, it has been published~\cite{bhargav_joshi_2022_7510572} on Zenodo\footnote{\href{https://zenodo.org/}{https://zenodo.org/}} platform, which was launched in May 2013 as part of the OpenAIRE project, in partnership with CERN. 
The dataset consists of 26 files in \textit{tar gzip} format, each file consisting of up to 10 csv files that contain measurements of up to 360 crystals. The files corresponding to the +z side of the ECAL are labelled as "plus" and those corresponding to -z side are labelled "minus". The list of $i\eta$ rings along with their position in terms of pseudo-rapidity ($\eta$) and azimuthal angle ($\phi$) in each tar file are included in form of json files under the metadata section in Zenodo.

\section{Machine Learning Solutions}

We describe in this and subsequent sections two ML solutions that we have developed to address this problem, their training and the results we have obtained. The source code is published on GitHub at \textcolor{blue}{\href{https://github.com/FAIR-UMN/fair_ecal_monitoring}{https://github.com/FAIR-UMN/fair\_ecal\_monitoring}}.

Neural networks (NNs) are machine learning (ML) methods that mimic the biological structure and functioning of neurons in a brain. A NN that does not involve any cyclic connections is called a Feedforward neural network (FNN). Deep neural networks (DNNs) are structures that consist of many stages of interconnected neurons. DNNs perform well for hard learning tasks, such as object identification and speech recognition. However, they require that the inputs and outputs of the task be encoded into vectors with fixed dimensionality \cite{sutskever2014sequence}. 


Many problems, like machine translation and speech recognition, have a sequential structure, since their input and output lengths are not known \textit{a-priori}~\cite{sutskever2014sequence}. 
A class of neural networks called recurrent neural networks (RNNs) are a type of FNNs that pass the data sequentially between different nodes. This architecture allows the network to learn and to retain past knowledge when processing data points from a given data series. RNNs have some shortcomings---in particular, the vanishing gradient problem~\cite{vanishing-gradient} while training the network. Other architectures have been developed in the past to address these issues.

\subsection{Long Short Term Memory (LSTM) Models}

LSTM models are a type of RNNs that include feedback components. RNNs are good at tracking arbitrary long-term dependencies in a sequence but have a tendency to be unstable during training. LSTMs solve the vanishing-gradient problem through an additive gradient structure. The LSTM cell is shown in Figure \ref{fig:encoder-decoder}, which is the key part of the Seq2Seq model. For each element in the input sequence, each layer of the LSTM computes the following functions: 
\begin{align}
    \begin{split}
        & \vi_t = \sigma(\mW_{ii}\vx_t + \mathbf{b}_{ii} + \mW_{hi}\vh_{t-1} + \mathbf{b}_{hi} ) \\
        & \vf_t = \sigma(\mW_{if}\vx_t + \mathbf{b}_{if} + \mW_{hf}\vh_{t-1} + \mathbf{b}_{hf} ) \\ 
        & \vg_t = \tanh(\mW_{ig}\vx_t + \mathbf{b}_{ig} + \mW_{hg}\vh_{t-1} + \mathbf{b}_{hg} ) \\
        & \vo_t = \sigma(\mW_{io}\vx_t + \mathbf{b}_{io} + \mW_{ho}\vh_{t-1} + \mathbf{b}_{ho} ) \\
        & \vc_t = \vf_t \odot \vc_{t-1} + \vi_t \odot \vg_t \\
        & \vh_t = \vo_t \odot \tanh{\vc_t} 
    \end{split}
\end{align}

where $\vh_t,\vc_t$ are the hidden and cell state at time $t$, $\vx_t$ is the input at time $t$, $\vi_t,\vf_t,\vg_t,\vo_t$ are the input, forget, cell, and output gates, respectively. $\sigma$ is the sigmoid function and $\odot$ is the Hadamard product~\cite{lstm}.


\subsection{Seq2Seq Model}

A sequence-to-sequence (Seq2Seq) model is an architecture that combines two or more LSTMs. It consists of two parts---the encoder and the decoder, each of which is built by using separate LSTMs. This type of model has been developed for automatic language translation, where a sentence from one language is translated to another language \cite{sutskever2014sequence}. The encoder is used to process each token in the input sentence, and encode all the input sequence information into a fixed-length vector. The transform vector, known as context vector, is a vector in a latent space and it encapsulates the whole meaning of the input sequence. The decoder reads the context vector and predicts the target sequence token by token.

\begin{figure}[!ht]
  \centering
  \includegraphics[width=1.0\textwidth]{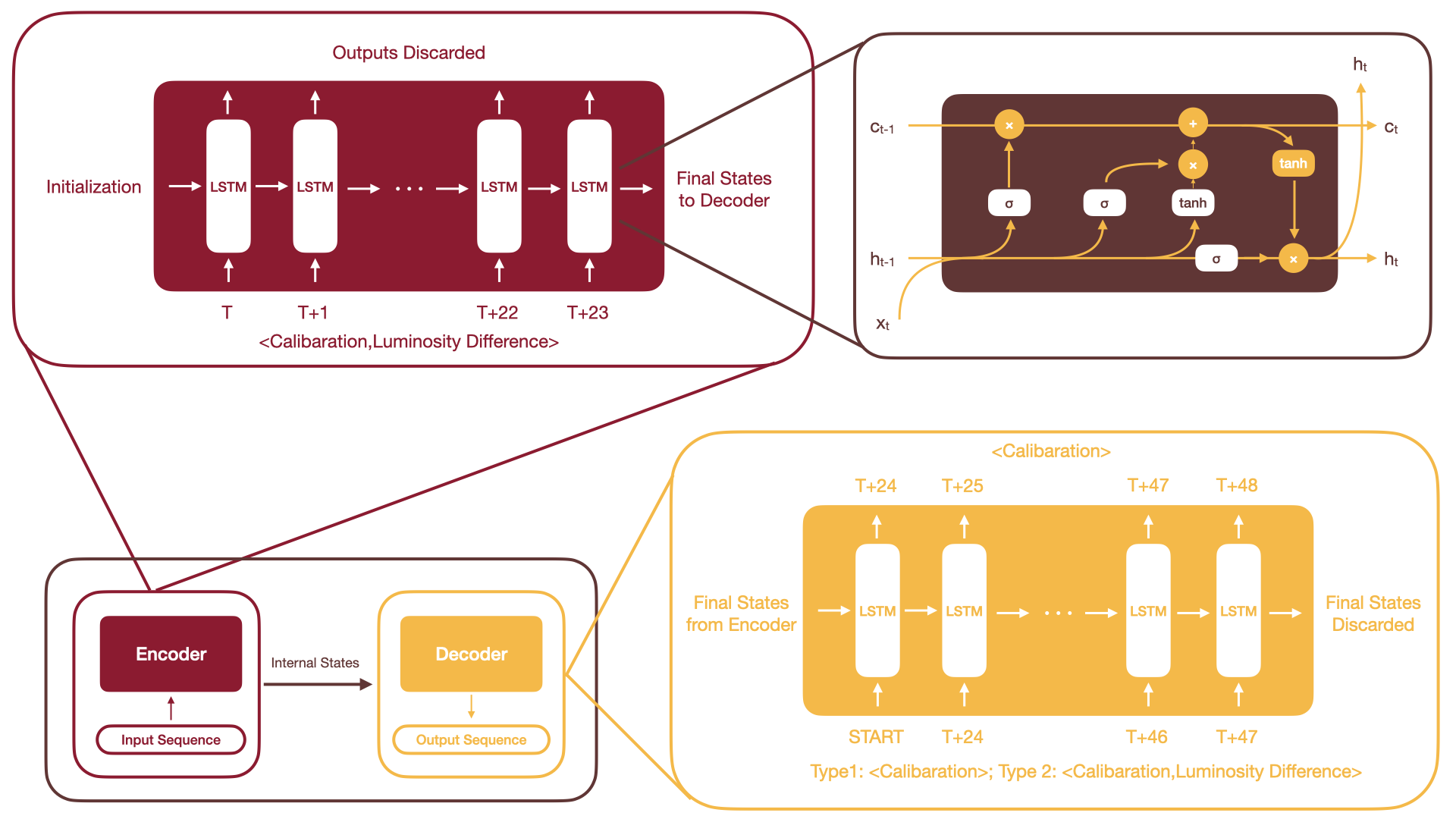}
  \caption{Seq2Seq model for used for predicting future calibration values (\textbf{lower left}). The Encoder block (\textbf{upper left}) and Decoder block (\textbf{lower right}) are a set of sequentially connected LSTM units (\textbf{upper right}).}
  \label{fig:encoder-decoder}
\end{figure}

Figure \ref{fig:encoder-decoder} shows the basic architecture of the encoder-decoder network used for this problem. The encoder block consists of LSTM units connected in series that take in a set of calibration values and the luminosity differences between those calibration values as input. All the information from the input sequence is encapsulated in the internal states $\vh_t$ (hidden state) and $\vc_t$ (cell state). The decoder block is another block of LSTM units connected in series. The final states $(\vh_t,\vc_t)$ of the encoder are used as the initial states $(\vh_0,\vc_0)$ of the decoder, which is the context vector used to predict the target sequence. The decoder network also takes input along with the initial states to predict the target sequence. The input to the decoder varies according to the method used for the training.

The Seq2Seq model can be trained using teacher forcing method, where the decoder is trained using the target output (ground truth output) instead of the output generated by the decoder in the previous step of the sequence. However, during the evaluation step, the decoder generates the output sequentially using the output generated in the previous step. Using teacher forcing during training has been shown to improve the training process. To determine if the training process has converged, the model was trained for 300 epochs and the validation e

\section{Training}

For training AI models, PyTorch packages were used. The code is maintained in a public Github repository (\url{https://github.com/FAIR-UMN/FAIR-UMN-ECAL}). Conda environments are provided so that the users of the datasets can use any API of their liking. The project details can also be found in \url{https://fair-umn.github.io/FAIR-UMN-Docs}.


\subsection{Data Pre-Processing}


1080 csv files from the datasets mentioned in \cref{datasets} are used for the training and prediction. 360 csv files (\texttt{df\_skimmed\_xtal\_54000\_2016} to \texttt{df\_skimmed\_xtal\_54359\_2016}; size 203 MB) are used for training, and 720 csv files (\texttt{df\_skimmed\_xtal\_54000\_2017} to \texttt{df\_skimmed\_xtal\_}\\\texttt{54359\_2017} and \texttt{df\_skimmed\_xtal\_54000\_2018} to \texttt{df\_skimmed\_xtal\_54359\_2018}; size 406 MB) are used for prediction. 

The difference between subsequent entries in the dataset is the integrated luminosity delivered between the two consecutive measurements. Before training the networks, the measured calibration values and the luminosity differences in the training dataset are normalized to unity using the \textit{StandardScaler} from the sklearn library. Next, to obtain the input $X$ and the true output $Y_{\text{true}}$ used for model training, we performed the following steps:
\begin{itemize}
\item Define the input length $L_E$ (e.g., $L_E=24$), corresponding to the number of LSTM units in the encoder, and the output length $L_D$ (e.g., $L_D=24$), corresponding to the number of LSTM units in the decoder for each individual sample. The Seq2Seq model will be trained to use a sequence of calibration values and luminosity differences of length $L_E$ and learn to predict the next $L_D$ calibration values. 
\item To avoid any overlap between the prediction sequences, a separation stride $L_S$ is set to be the same as $L_{D}$. Therefore, the total number of samples is 
\[N_{sample} = \frac{N-L_{E}-L_{D}}{L_{S}}+1,\]
where N is the total number of entries in the training dataset. For each individual sample, the input is a sequence starting from $T$ to $T+L_{E}-1$ and the output is a sequence starting from $T+L_{E}$ to $T+L_{E}+L_{D}-1$. 
\item In PyTorch, the LSTM module takes a 3D tensor as the input whose dimensions are given by (sequence, batch, features). In this problem, the input to encoder (X$_{\text{encoder}}$), the input to decoder (X$_{\text{decoder}}$) and the output of the decoder (Y$_{\text{decoder}}$) can be represented as below:
\begin{align*}
    &X_{\text{encoder}} \in \R^{L_{E}\times N_{\text{sample}}\times N_{E}},\\
    &X_{\text{decoder}} \in \R^{L_{D}\times N_{\text{sample}}\times N_{D}},\\
    &Y_{\text{decoder}} \in \R^{L_{D}\times N_{\text{sample}}\times 1},
\end{align*}
where N$_{E}$ and N$_{D}$ are the number of features used in the encoder and decoder, respectively. In this study, N$_{E}$ and N$_{D}$ are set to 2, which represents features calibration value and luminosity difference. But more features, such as the difference in the timestamps between two entries, can be added if needed.
\end{itemize}

\subsection{Training Seq2Seq Model}
The Seq2Seq model was built using the PyTorch library. Both encoder and decoder blocks were set with 1024 hidden layers. The number of LSTM cells in the encoder and the decoder is varied to scan for the optimal input and output lengths. The LSTMs were initialized with a sigmoid activation for input, forget, and output gates and a hyper-tangent activation for the cell gates. The Mean Squared Error (MSE) loss function along with the Adam~\cite{kingma2014adam} optimizer is used to train the model. The model is trained for 200 epochs with a batch size of 128 and a learning rate of $10^{-3}$. A higher number of epochs (3000) were used to check if the model performance improves, but it was found that it converges after about 200 epochs. All trainings and predictions were performed on machine with Intel Xeon Silver 4214R@2.40GHz, and Nvidia RTX A6000 graphics card with 48 GB memory.

\begin{figure}[ht]
  \centering
  \includegraphics[width=0.45\textwidth]{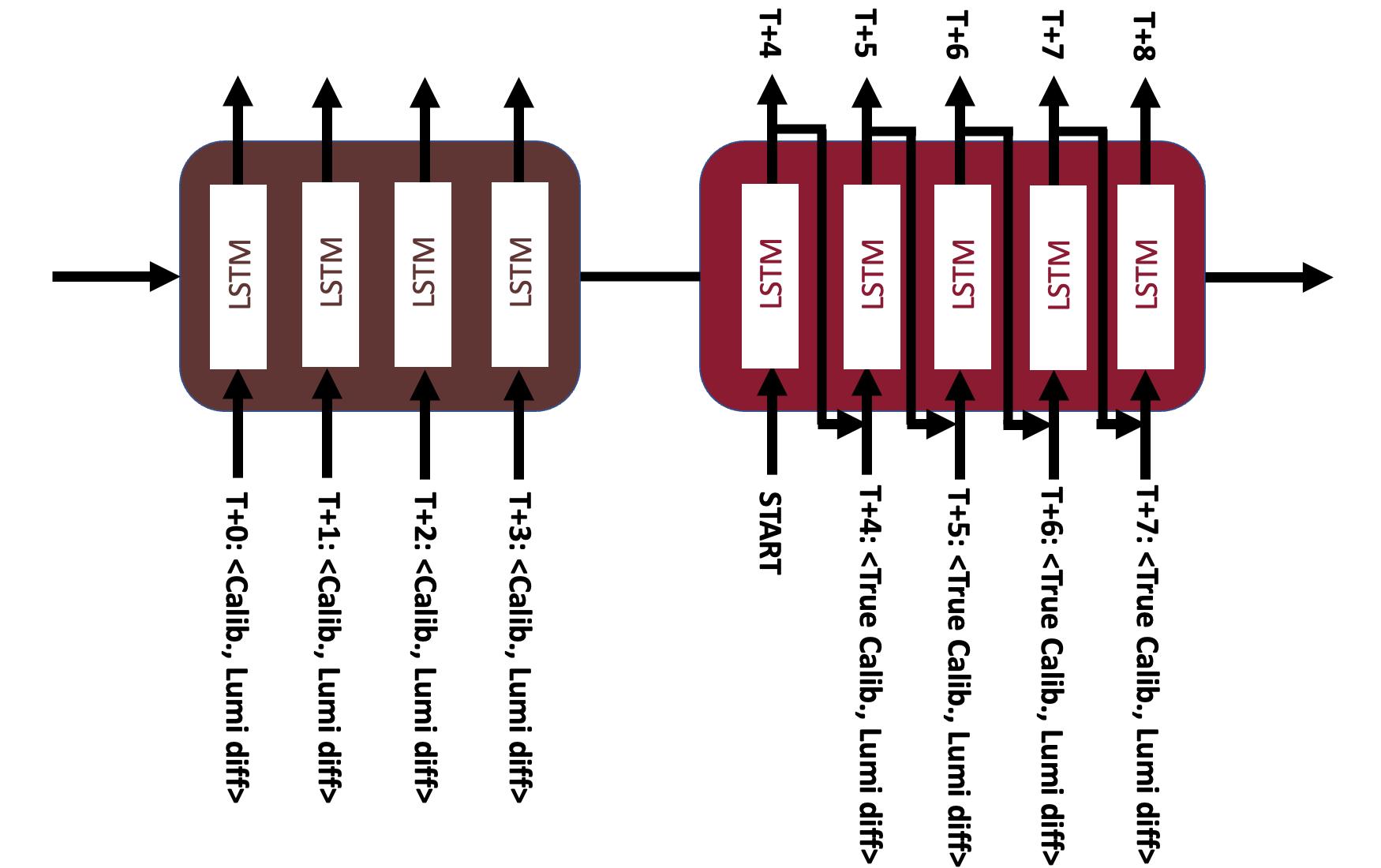}
  \includegraphics[width=0.45\textwidth]{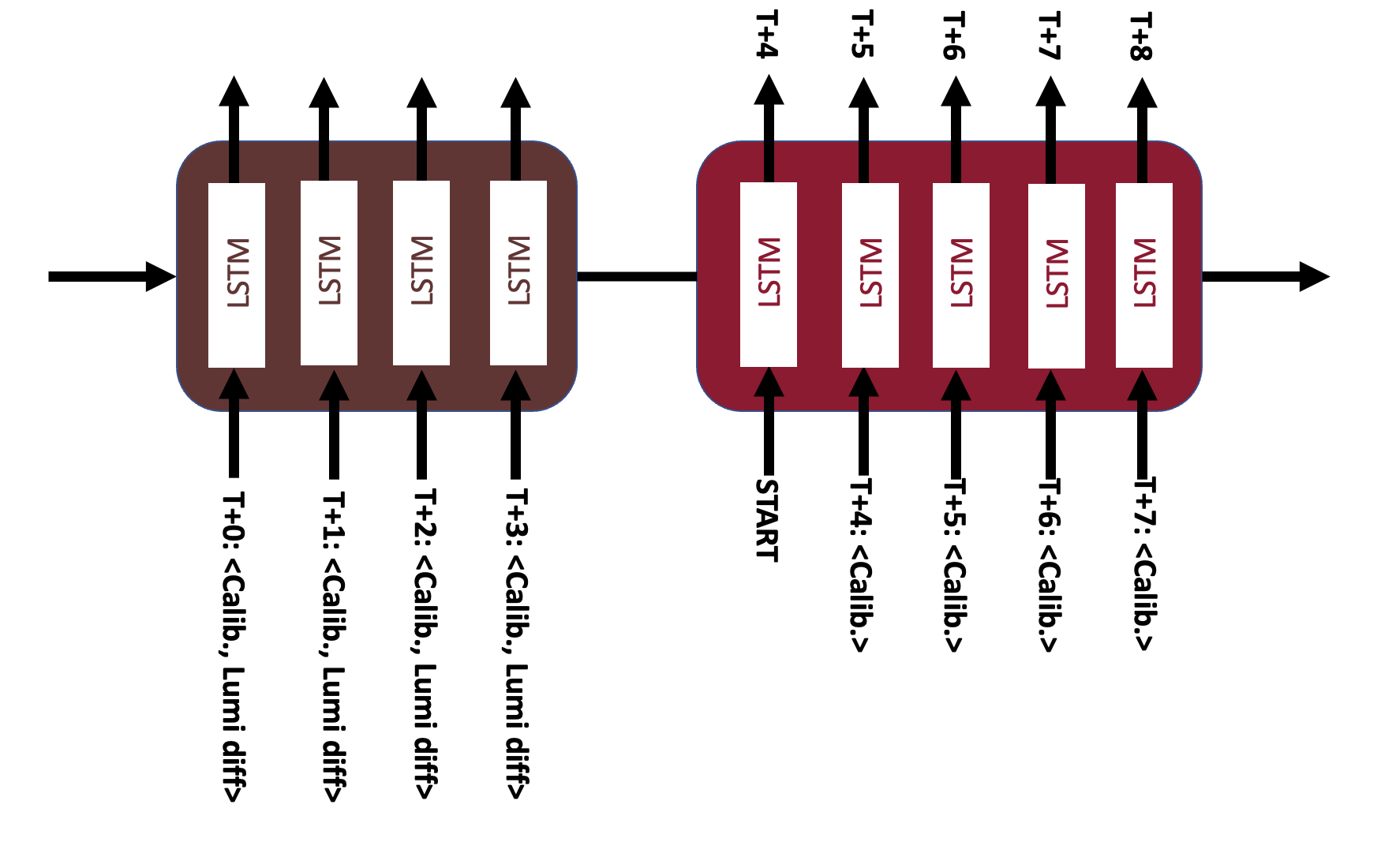}
  \caption{\textbf{(left)}: Seq2Seq model with recursive training; \textbf{(right)}: Se2Seq model with teacher forcing.}
  \label{fig:train_strg}
\end{figure}

With a large number of data points, there are several options available for training a model. A single model can be developed for each of the individual crystals. On the other hand, the data points of different crystals that are at equal distances from the center of ECAL, i.e., within one i$\eta$-ring of the ECAL, can be combined together. The assumption is that these crystals receive an equal amount of radiation dose because of the radial symmetry and hence will have similar behavior in laser response over a course of time. Then this model trained with more data points would be able to predict calibrations for all the crystals in the corresponding i$\eta$-ring. In addition to changing the number of crystals, three different strategies were used which are given as follows:
\begin{enumerate}
\item Recursive: In this setting, as shown in Figure \ref{fig:train_strg} (left), we feed the token from $Y_{\text{pred}}$ from the previous time step as the input to the current time step.
\item Teacher Forcing: In this setting, as shown in Figure \ref{fig:train_strg} (right), we feed the token from $Y_{\text{true}}$ (instead of the token from $Y_{\text{pred}}$) from the previous time step as the input to the current time step.
\item Mixed: In this setting, the previous two strategies can be combined in different ratios. For example, a mixed training with a teacher forcing ratio of 0.7 means that only 70\% of the batches in the decoder training use the teacher forcing strategy. 
\end{enumerate}

\section{Results}
To quantify the performance of a LSTM or Seq2Seq model, we use Mean Absolute Percentage Error (MAPE) as the metric, which is defined as
\begin{align}
    \text{MAPE} = \frac{100\%}{n} \sum_{t-1}^n \abs{\frac{A_t-F_t}{A_t}},
\end{align}
where $A_t$ is the actual value and $F_t$ is the forecast value.

\begin{figure}[!ht]
  \centering
  \includegraphics[width=0.75\textwidth]{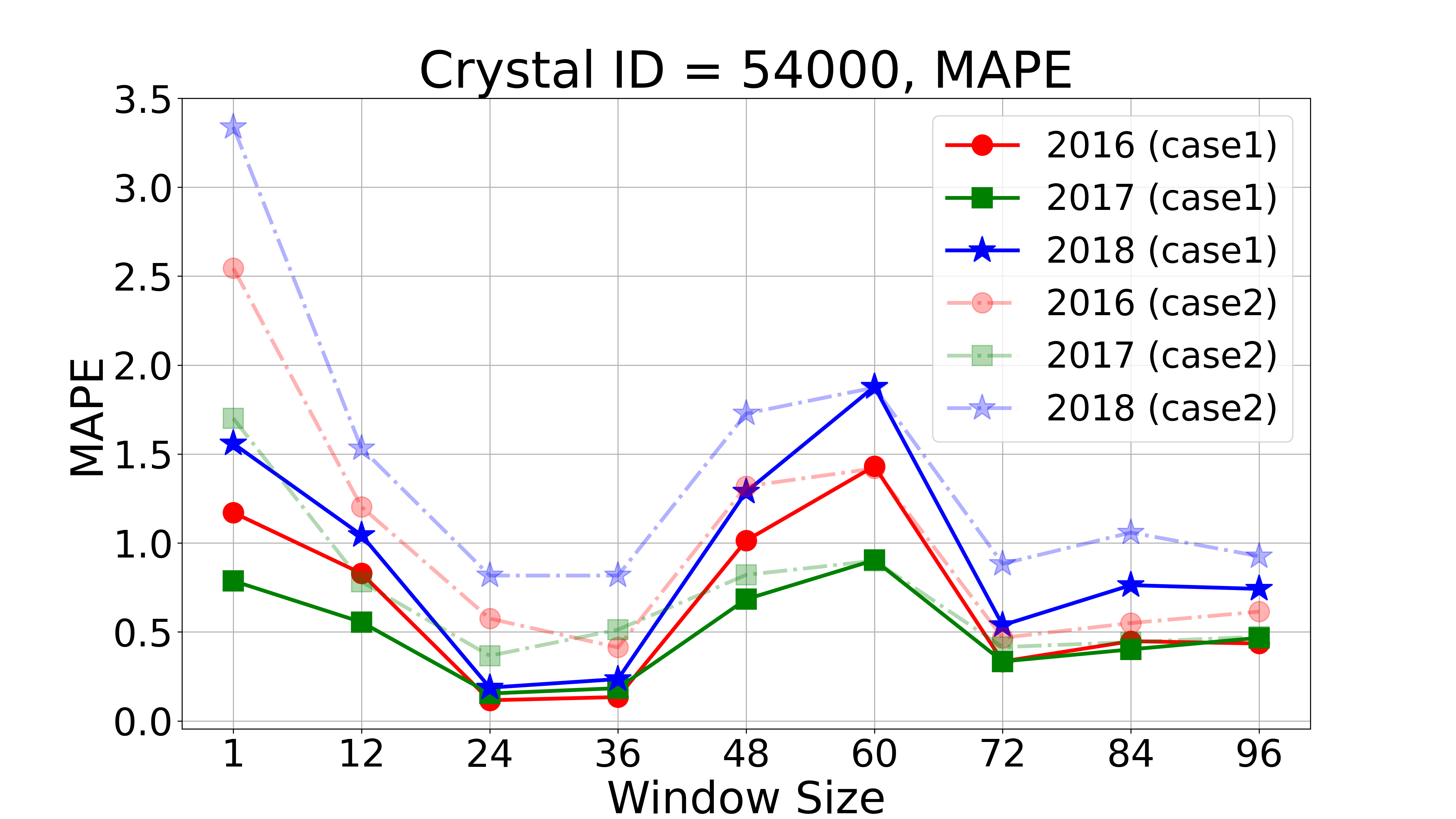}

  \caption{Seq2Seq model trained with different sequence lengths.} 
  \label{fig:diff_window_size}
  \end{figure}

To determine an appropriate input length ($L_E$) and output length ($L_D$) for the Seq2Seq model, the models were trained using different sequence lengths with the recursive training strategy. We set $L_E=L_D$ in this example. As shown in Figure \ref{fig:diff_window_size}, the sequence length equals $24$ gives the lowest MAPE. Hence, this value will be used for all the trainings described in this section.

\subsection{Different Training Strategies}

For the purpose of this study, the data taken in the year 2016 corresponding to single crystal with ID 54000, from i$\eta$-ring 66, were used to train a Seq2Seq model (Model-S). Another model (Model-R) was trained using the data taken in the year 2016 corresponding to all 360 crystals in i$\eta$-ring 66, with ID ranging from 54000 to 54359. The response of the two models was evaluated on different crystal from the same i$\eta$-ring 66 using the data taken in the year of 2017 and 2018.

\begin{figure}[!ht]
  \centering
  \includegraphics[width=0.5\textwidth]{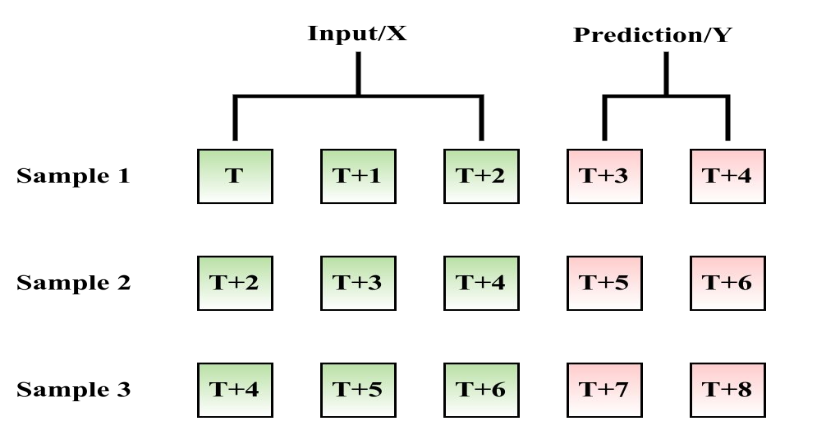}
  \includegraphics[width=0.48\textwidth]{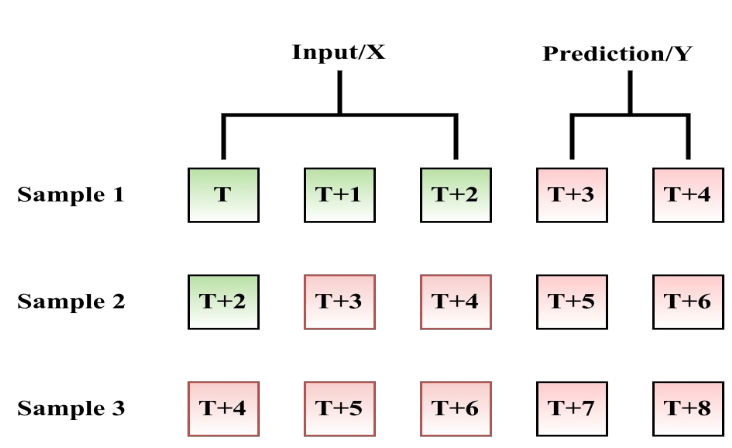}
  \caption{\textbf{(left)}: Case 1: without using prediction as the input of the next round prediction; \textbf{(right)}: Case 2: using prediction as the input of the next round prediction}
  \label{fig:2cases}
\end{figure}

For making these predictions, two cases were used:
\begin{enumerate}
\item Case 1: The ground truth is provided as the model input at each prediction window (Figure \ref{fig:2cases} (left)).
\item Case 2: In this case only the first input is provided and the model would recursively ``reuses" the predictions from its previous prediction window as its input (Figure \ref{fig:2cases} (right)).  to make predictions and then evaluate their performance separately. Therefore, in terms of learning, this case is more challenging than Case 1 as we use less prior information.
\end{enumerate}

The input to each LSTM cell in the decoder contains the luminosity delivered ($\Delta L_i$) between the current and the next timestamp, and the previous calibration value. Typically, during training, the output of the previous LSTM cell in the decoder is used as input to the next LSTM as a current calibration value, along with the $\Delta L_i$. However, in case of teacher forcing, the true calibration value for input is used instead of the output from the previous LSTM cell. The teacher forcing ratio can be used to define the fraction of batches in the decoder training that would get true calibration values as input during training.

\begin{figure}[!ht]
  \centering
  \includegraphics[width=0.495\textwidth]{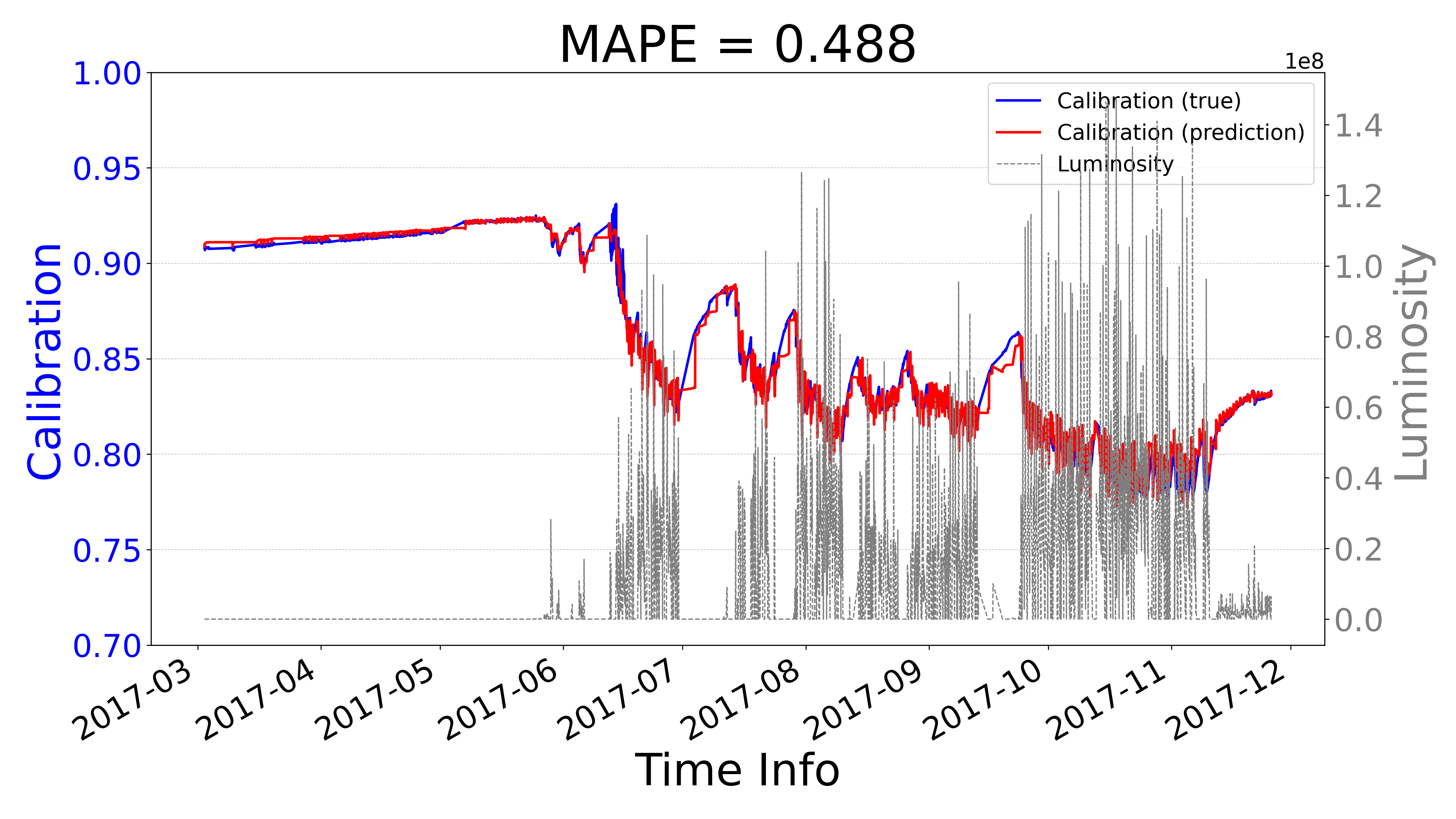}
  \includegraphics[width=0.495\textwidth]{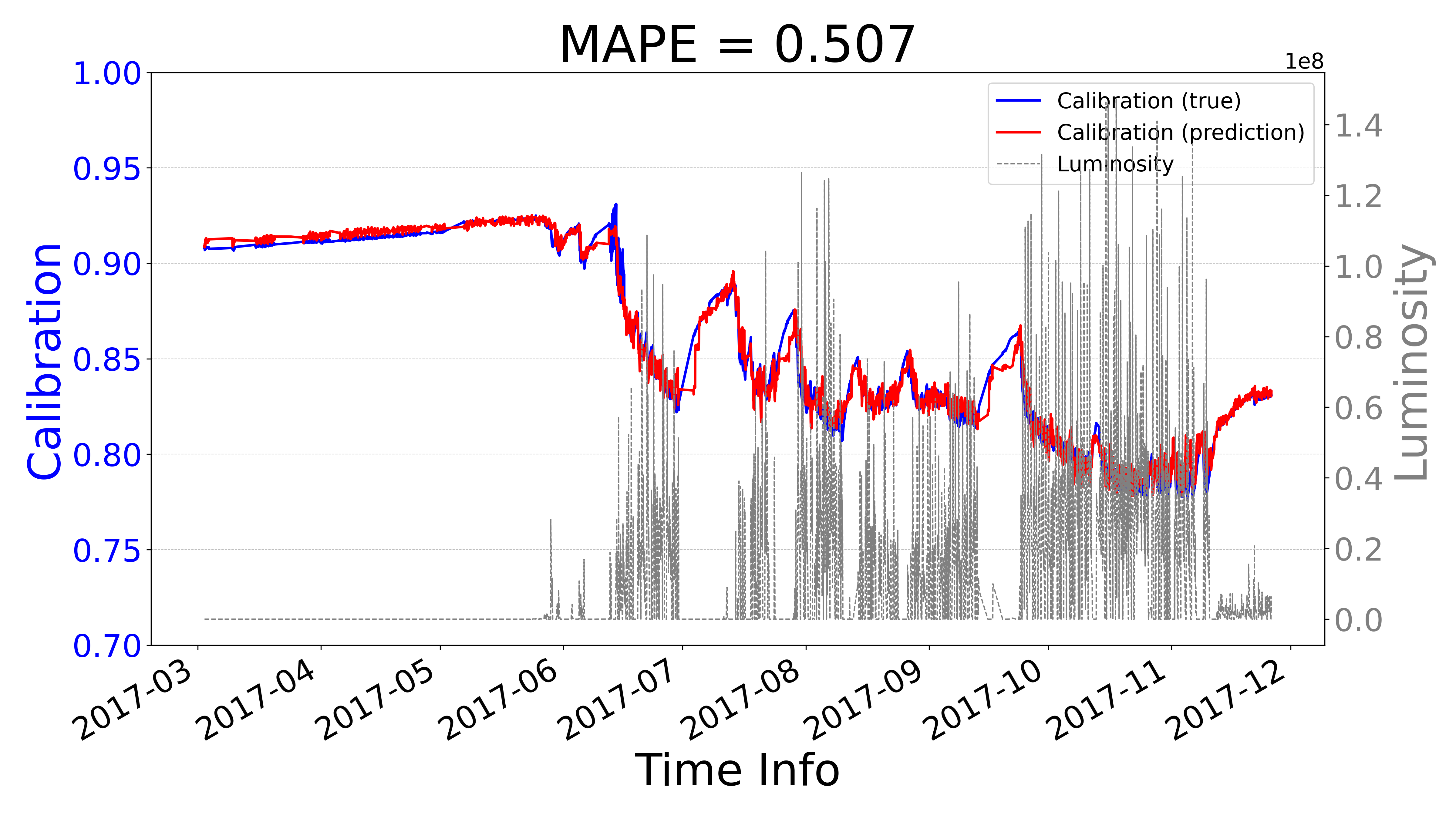}\\
  \includegraphics[width=0.495\textwidth]{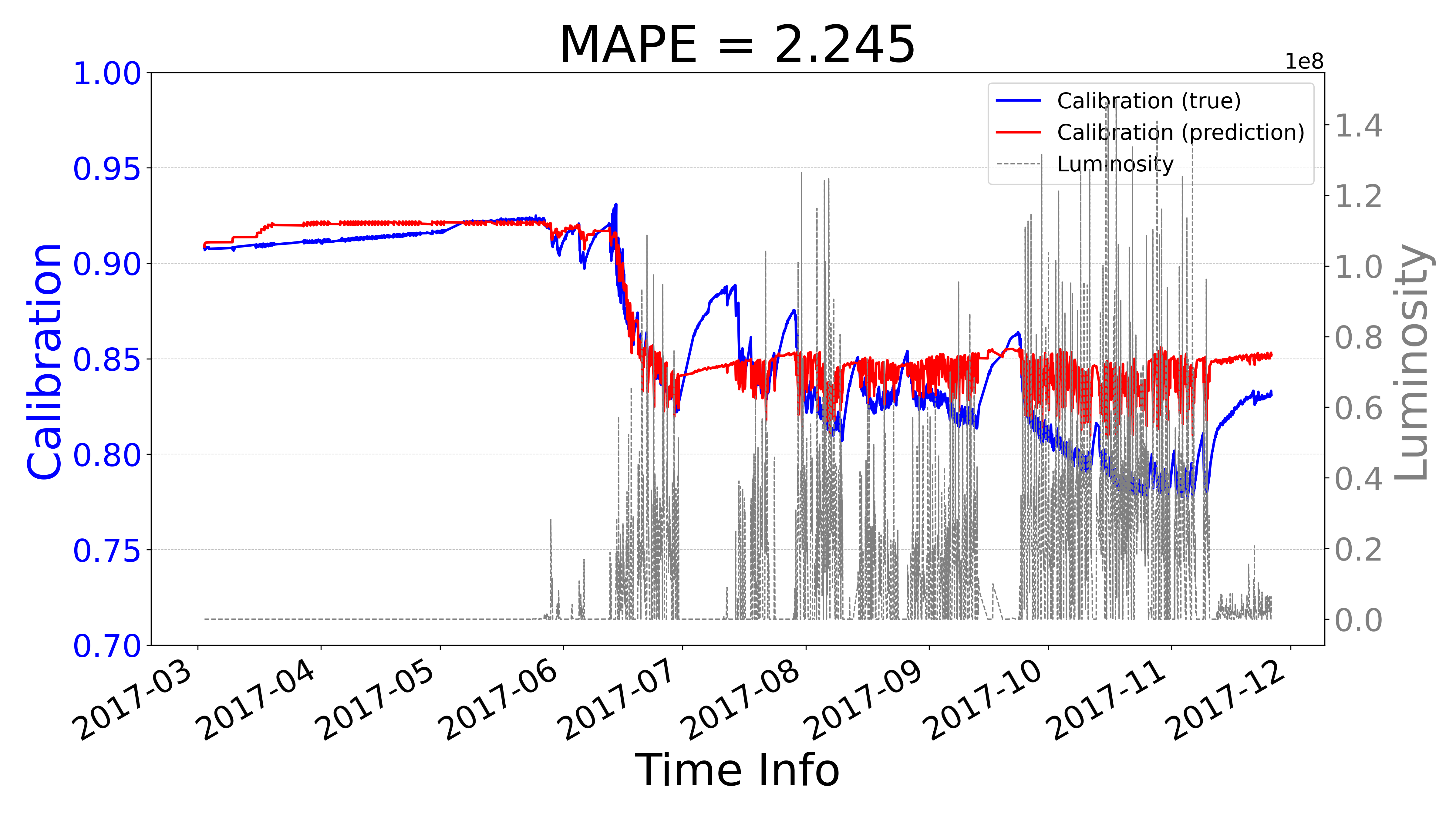}
  \includegraphics[width=0.495\textwidth]{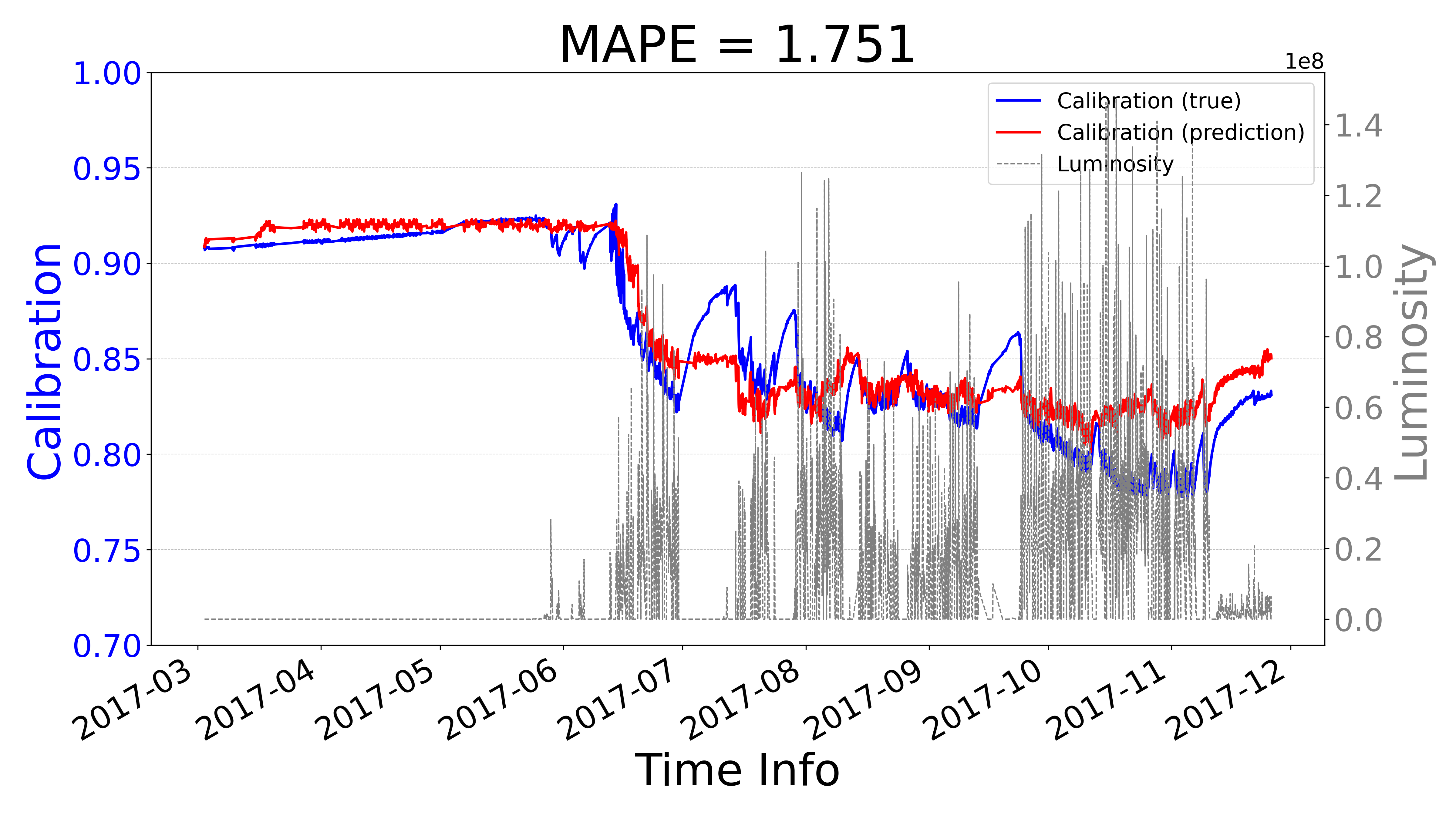}
  \caption{A demonstration of the true calibration curve vs the predicted calibration curve. (\textbf{top left}) Case 1 predictions using Model-S (mixed; teacher forcing ratio = 0.5): the response of the crystal ID 54300 in 2017. (\textbf{top right}) Case 1 predictions using Model-R (mixed; teacher forcing ratio = 0.5): the response of the crystal ID 54300 in 2017. (\textbf{bottom left}) Case 2 predictions using Model-S (mixed; teacher forcing ratio = 0.5): the response of the crystal ID 54300 in 2017. (\textbf{top right}) Case 2 predictions using Model-R (mixed; teacher forcing ratio = 0.5): the response of the crystal ID 54300 in 2017.} 
  \label{fig:demo}
\end{figure}

The demonstration of our predicted calibration curve is shown in Figure \ref{fig:demo}: Both Model-S and Model-R in Case 1 can successfully predict the calibration in future time steps with low MAPE. However, both Model-S and Model-R in Case 2 give worse predictions after certain time steps.



\begin{figure}[!ht]
  \centering
  \includegraphics[width=0.995\textwidth]{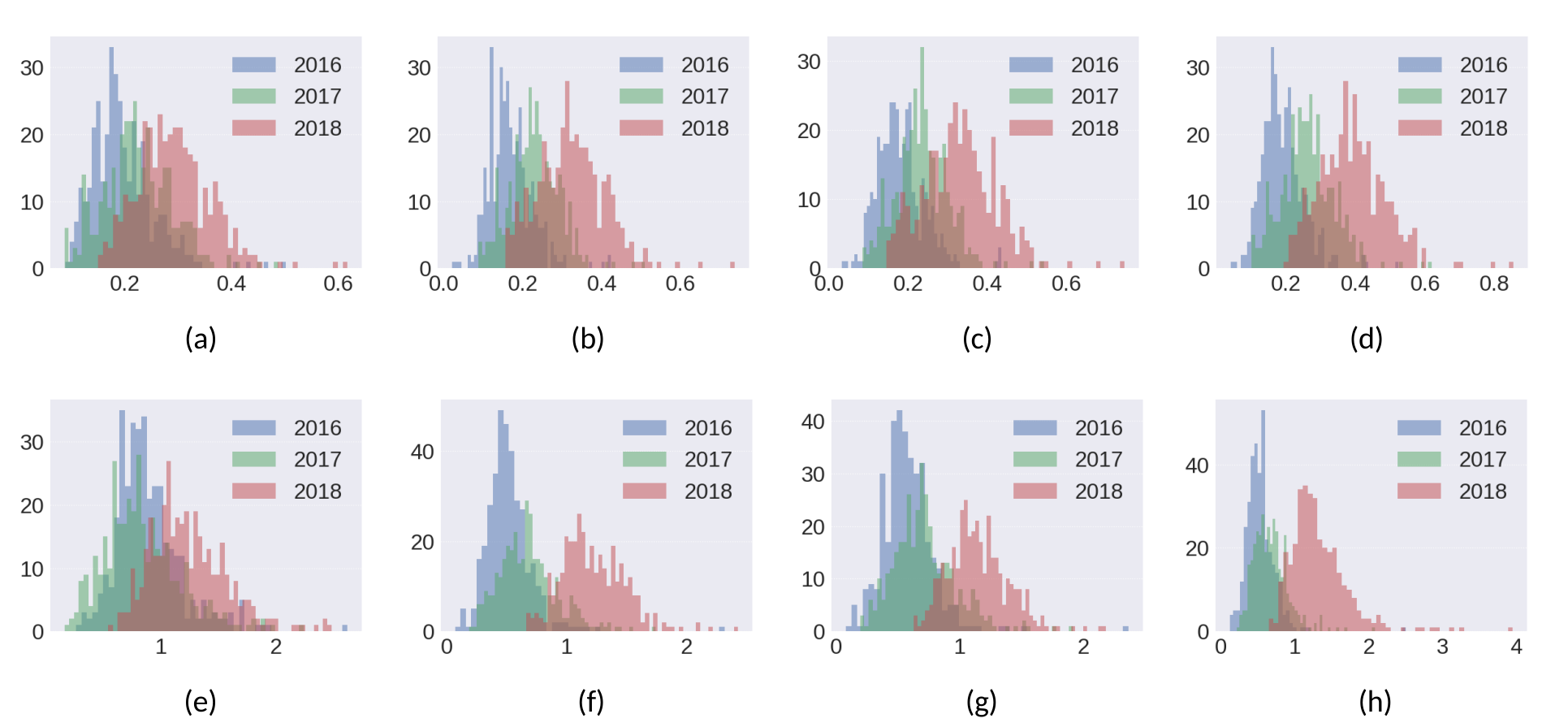}
  \caption{MAPE histograms from prediction on year 2016, 2017 and 2018. (a)(b)(c)(d) use Case 1 prediction strategy, while (e)(f)(g)(h) use Case 2 prediction strategy. (a) and (e) use model trained on single crystal 54000 of 2016. (b)(c)(d)(f)(g)(h) use model trained on all crystals in ring 66 (crystal ID: 54000-54359) of 2016. (b)(f) use recursive strategy in the encoder; (c)(g) use mixed strategy (teacher forcing ratio $=0.5$) in the encoder; (d)(h) use teacher forcing strategy in the encoder.}
  \label{fig:multiple_mape_hist}
\end{figure}

\begin{table}[!ht]
\begin{center}
\begin{tabular}[t]{ |c|c|c|c|c|c| } 
 \hline
 Year & Prediction  & Model-S(M) & Model-R(R) & Model-R(M) & Model-R(T)  \\ 
 \hline
 2016 & Case 1 & 0.194 & 0.168 & 0.180 & 0.191 \\  
 2017 & Case 1 & 0.223 & 0.228 & 0.234 & 0.263 \\ 
 2018 & Case 1 & 0.291 &  0.323& 0.330 & 0.391 \\ 
 \hline
 2016 & Case 2 & 0.888 &  0.516& 0.577 & 0.530 \\  
 2017 & Case 2 & 0.836 &  0.680& 0.713 & 0.673 \\ 
 2018 & Case 2 & 1.24 &  1.216 & 1.147 & 1.327 \\ 
 \hline
\end{tabular}
\end{center}
\caption{Average MAPE from prediction on all 360 crystals. M: mixed strategy of teacher forcing and recursive (teacher forcing ratio$=0.5$); R: recursive strategy; T: teacher forcing strategy.}
\label{tab:MAPE}
\end{table} 

Furthermore, MAPE has been evaluated for all 360 crystals (crystal ID:34000-34359) using the data for three years (2016, 2017 and 2018) and the distribution is shown in Figure \ref{fig:multiple_mape_hist}. Also, the corresponding average MAPE among all predictions is shown in Table \ref{tab:MAPE}. As shown in both Figure \ref{fig:multiple_mape_hist} and Table \ref{tab:MAPE} Model-R with recursive, mixed, and teacher forcing strategy have different behavior: the recursive version leads to lower MAPE than the teacher forcing version, which indicates a potential overfitting when using the teacher forcing strategy. Also, in the Case 1 prediction, the mixed strategy has MAPE between recursive and teacher-forcing; while in the Case 2, the mixed strategy gets a worse prediction in the 2016 and 2017 data, but gets a better prediction in the 2018 data.

As shown in both Figure \ref{fig:multiple_mape_hist} (a)(c)(e)(g) and Table \ref{tab:MAPE}, Model-S (M) is worse than Model-R (M) in the Case 2 prediction but better than Model-R (M) in the Case 1 prediction (2017 and 2018). Besides that, the Case 1 prediction would be recommended, as it always has better performance than the Case 2 prediction. 



\section{Summary}

Recurrent neural networks, such as long short-term memory and sequence-to-sequence models, can be used to predict the response of the CMS electromagnetic calorimeter crystals in the future. The reduction in the crystal transparency as a function of the luminosity of the CERN LHC, or their recovery, when there is no radiation, can be captured by these models. The dataset of the time development of the response to laser light for each crystal and of the luminosity of the LHC, between 2016 and 2018 is now available publicly. We have demonstrated how standard machine learning tools can be used to predict future changes to the crystals.  It is our intent that by publishing our data and our models, including the source code, it will generate interest in the wider computer science and physics communities leading to the development of novel tools to address the problem of how to predict the future performance of the CMS crystals in the both the short term,  and as far ahead as 2035, when the CMS detector will still be taking data.

\section*{Acknowledgments}
This work has been supported by the Department of Energy, Office of Science, Office of Advanced Scientific Computing under award number DE-SC0021395. 
The authors would like to express their gratitude to the CMS Collaboration, and in particular to the CMS ECAL community for making the crystal transparency data available, and to the CMS beam, radiation and integrated luminosity group for supplying the luminosity data. We would also like to thank our colleagues from the FAIR4HEP group for discussions and their invaluable inputs and suggestions for writing this paper.

\bibliographystyle{unsrt}  
\bibliography{references}

\begin{thebibliography}{10}

\bibitem{fairmetrics}
Mark~D. Wilkinson, Susanna-Assunta Sansone, Erik Schultes, Peter Doorn, Luiz
  Olavo~Bonino da~Silva~Santos, and Michel Dumontier.
\newblock A design framework and exemplar metrics for {FAIRness}.
\newblock {\em Scientific Data}, 5(1):180118, May 2018.

\bibitem{CMS_2008}
The~CMS Collaboration.
\newblock The {CMS} experiment at the {CERN} {LHC}.
\newblock {\em Journal of Instrumentation}, 3(08):S08004--S08004, aug 2008.

\bibitem{Chatrchyan:2012ufa}
Serguei Chatrchyan et~al.
\newblock Observation of a new boson at a mass of 125 {GeV} with the {CMS}
  experiment at the {LHC}.
\newblock {\em Phys. Lett. B}, 716:30, 2012.

\bibitem{HGG}
A.~M. Sirunyan et~al.
\newblock Measurements of higgs boson production cross sections and couplings
  in the diphoton decay channel at $\sqrt{s}$ = 13 tev.
\newblock {\em Journal of High Energy Physics}, 2021(7):27, Jul 2021.

\bibitem{Chen_2022}
Yifan Chen, E.~A. Huerta, Javier Duarte, Philip Harris, Daniel~S. Katz, Mark~S.
  Neubauer, Daniel Diaz, Farouk Mokhtar, Raghav Kansal, Sang~Eon Park,
  Volodymyr~V. Kindratenko, Zhizhen Zhao, and Roger Rusack.
\newblock A {FAIR} and {AI}-ready higgs boson decay dataset.
\newblock {\em Scientific Data}, 9(1), feb 2022.

\bibitem{ECALTDR}
{\em The CMS electromagnetic calorimeter project: Technical Design Report}.
\newblock Technical design report. CMS. CERN, Geneva, 1997.

\bibitem{914439-xtal-color-centers}
Xiangdong Qu, Liyuan Zhang, and Ren-Yuan Zhu.
\newblock Radiation induced color centers and light monitoring for lead
  tungstate crystals.
\newblock {\em IEEE Transactions on Nuclear Science}, 47(6):1741--1747, 2000.

\bibitem{bhargav_joshi_2022_7510572}
Bhargav Joshi and Roger Rusack.
\newblock Laser response in {ECAL} crystals in {CMS} detector, March 2022.

\bibitem{sutskever2014sequence}
Ilya Sutskever, Oriol Vinyals, and Quoc~V Le.
\newblock Sequence to sequence learning with neural networks.
\newblock {\em Advances in neural information processing systems}, 27, 2014.

\bibitem{vanishing-gradient}
Razvan Pascanu, Tomas Mikolov, and Yoshua Bengio.
\newblock On the difficulty of training recurrent neural networks.
\newblock 2012.

\bibitem{lstm}
Lstm-pytorch 1.12 documentation.

\bibitem{kingma2014adam}
Diederik~P Kingma and Jimmy Ba.
\newblock Adam: A method for stochastic optimization.
\newblock {\em arXiv preprint arXiv:1412.6980}, 2014.

\end{thebibliography}

\end{document}